# Library and Culture: A Scientometric Analysis and Visualization of Research Trends

Auwalu Abdullahi Umar[1*], Muneer Ahmad[2], Dr M Sadik Batcha[3]

[1]*MLIS Student, Department of Library and Information Science, Annamalai University, Annamalai nagar, India*

[2]*Ph.D Research Scholar, Department of Library and Information Science, Annamalai University, Annamalai nagar, India*

[3]*Professor and University Librarian, Annamalai University, Annamalai nagar, India*

***Corresponding Author:*** *Auwalu Abdullahi, Umar MLIS Student, Department of Library and Information Science, Annamalai University, Annamalai nagar, India*

**ABSTRACT**

*The significance of libraries in preserving and maintaining history and traditional culture cannot be overlooked. It is from this purpose that libraries are to envisage in their programmes cultural activities which must be collected, documented and preserved for posterity. The usefulness of preserved information lies in the fact that the generation to come will be able to establish their identity. This will also assist them with a foundation to build from. This study focus on the growth and development of Library and Culture research in forms of publications reflected in Web of Science database, during the span of 2010-2019. A total 890 publications were found and the highest 124 (13.93%) publications published in 2019.The analysis maps comprehensively the parameters of total output, growth of output, authorship, institution wise and country-level collaboration patterns, major contributors (individuals, top publication sources, institutions, and countries). It exposed that the most prolific author is Lo P secured first place by contributing 4 (0.45%) publications, followed by Bressan V 3 (0.34%) publications in Library and Culture literature. Journal of Academic Librarianship produced the highest number of records 29 (3.26%) followed by Australian Library Journal having contributed 21 (2.36%).It is identified the domination of Wuhan University; School Information Management had contributed 6 (0.67%) of total research output. Authors from USA published the highest number of publications with a total of 244 (27.42%), followed by UK and Australia with 118 (13.26%) and 76 (8.54%) publications were produced respectively.*

## INTRODUCTION

### Library

The word Library has been derived from the Latin word "Libraria" which means a place where books and other reading materials are stored. According to the Oxford English Dictionary "Library is a building, room or set of rooms, containing a collection of books for the use for the public or of some particular portion of it, or of the members of some society, or the like; a public institution or establishment, charged with the care of a collection of books, and the duty of rendering the books accessible to those who required to use them"(Oxford English Dictionary, 1933).

The above definition of library has undergone a significant change with the changing times, civilization and culture. The modem definition of a library is a place, where documents containing knowledge and information are stored technically and scientifically processed, properly preserved and made easily available to the users when warranted without loss of time. The library is also sometimes referred to as the "memory of human race". Library is a fountainhead of information and knowledge. It can be compared to a giant brain that remembers all that the scientists, the historians, the poets, title philosophers, and other great intellectual have thought and learned. In short a library is a place where the experience and expertise of the past can meet the needs of the present(Khanna, 1989).

Hence a Library can be defend as a collection of graphic acoustic and holistic material such as books, periodicals, newspapers, manuscripts, maps, charts, filmstrips, microfilms, photographs, records, or any recorded piece of information systematically arranged and a designed for use. It functions to collect organized and disseminate knowledge/information to users.

Library and Culture: A Scientometric Analysis and Visualization of Research Trends

### Culture

Culture is a modern concept based on a term first used in classical antiquity by the Roman orator Cicero: "cultura animi". The term "culture" appeared first in Europe in the 18th and 19th centuries, to connote a process of cultivation or improvement, as in agriculture or horticulture. In the 19th century, the term developed to refer first to the betterment or refinement of the individual, especially through education, and then to the fulfillment of national aspirations or ideals. In the mid-19th century, some scientists used the term "culture" to refer to a universal human capacity. For the German nonpositivist sociologist Georg Simmel, culture referred to "the cultivation of individuals through the agency of external forms which have been objectified in the course of history". In the 20th century, "culture" emerged as a central concept in anthropology, encompassing the range of human phenomena that cannot be attributed to genetic inheritance. Specifically, the term "culture" in American anthropology had two meanings: the evolved human capacity to classify and represent experiences with symbols, and to act imaginatively and creatively; and the distinct ways that people living differently classified and represented their experiences, and acted creatively.

### Library Culture

A library culture has had a lasting effect on archival infrastructures is in the world. Here the terms 'public archives tradition' and 'historical manuscripts tradition' have very specific meanings. These were originally derived from a perceived distinction in the way written materials should be defined and handled depending on where and how they were managed. If they were managed by a body whose main function was to service the archives generated by its employing body (e.g. the state) then they belonged to the public archive's tradition and in state archives. If they were managed by an institution whose remit was to collect materials from the world outside and to foster research then they belonged to the historical manuscript's tradition, and went to the state libraries. Libraries were the original bodies for taking in the latter type of material – which whether it was 'archival' or not was considered as 'manuscripts'. In-house State archival institutions developed later and many actually grew out of the manuscript rooms of State libraries.20 For example, the Minnesota State Archives, managing the records generated by the State, is just a department of the Minnesota Historical Society, which collects 'historical manuscripts' – even though in reality these may include the genuinely archival, organizational and, records etc.

Effective collaboration between libraries and their communities is critical as it paves way to smooth working relationships. If libraries have good relations with their community it will be easier for them to collect cultural information from them. There is also a great need to educate community on the importance and roles the library plays in preserving culture. If they understand, then they will give the information freely. Above all libraries are better placed in documenting and preserving culture for posterity. This is because information in them is easily accessible and they are open to everyone in the community. As they keep cultural information sources, community history will live long and also help to preserve the memory of the society being served.

### REVIEW OF LITERATURE

(Batcha & Ahmad, 2017) analysed Indian Journal of Information Sources and Services (IJISS) and Pakistan Journal of Library and Information Science (PJLIS) during 2011-2017 and studied various aspects like year wise distribution of papers, authorship pattern & author productivity, degree of collaboration pattern of Co-Authorship , average length of papers , average keywords, etc and found 138 (94.52%) of contributions from IJISS were made by Indian authors and similarly 94 (77.05) of contributions from PJLIS were done by Pakistani authors. Papers by Indian and Pakistani Authors with Foreign Collaboration are minimal (1.37% of articles) and (4.10% of articles) respectively.

(Ahmad, Batcha, Wani, Khan, & Jahina, 2018) explored scientometric analysis of the Webology Journal. The paper analyses the pattern of growth of the research output published in the journal, pattern of authorship, author productivity, and subjects covered to the papers over the period (2013-2017). It was found that 62 papers were published during the period of study (2013-2017). The maximum numbers of articles were collaborative in nature. The subject concentration of the journal noted was Social Networking/Web 2.0/Library 2.0 and Scientometrics or Bibliometrics. Iranian researchers contributed the maximum number of articles (37.10%). The study applied standard formula and statistical tools to bring out the factual results.

(Batcha, Jahina, & Ahmad, 2018) has examined scientometric analysis of the DESIDOC Journal

and analyzed the pattern of growth of the research output published in the journal, pattern of authorship, author productivity, and, subjects covered to the papers over the period (2013-2017). It found that 227 papers were published during the period of study (2001-2012). The maximum numbers of articles were collaborative in nature. The subject concentration of the journal noted was Scientometrics. The maximum numbers of articles (65 %) have ranged their thought contents between 6 and 10 pages.

(Ahmad & Batcha, 2019) analyzed research productivity in Journal of Documentation (JDoc) for a period of 30 years between 1989 and 2018. Web of Science database a service from Clarivate Analytics has been used to download citation and source data. Bibexcel and Histcite application software have been used to present the datasets. Analysis part focuses on the parameters like citation impact at local and global level, influential authors and their total output, ranking of contributing institutions and countries. In addition to this scientographical mapping of data is presented through graphs using VOSviewer software mapping technique.

(Ahmad & Batcha, 2019)studied the scholarly communication of Bharathiar University which is one of the vibrant universities in Tamil Nadu. The study finds out the impact of research produced, year-wise research output, citation impact at local and global level, prominent authors and their total output, top journals of publications, collaborating countries, most contributing departments and publication trends of the university during 2009 to 2018. The 10 years' publication data of the university indicate that a total of 3440 papers have been published from 2009 to 2018 receiving 38104 citations with h-index as 68. In addition, the study used sciento-graphical mapping of data and presented it through graphs using VOSviewer software mapping technique.

(Ahmad, Batcha, & Jahina, 2019) quantitatively identified the research productivity in the area of artificial intelligence at global level over the study period of ten years (2008-2017). The study identified the trends and characteristics of growth and collaboration pattern of artificial intelligence research output. Average growth rate of artificial intelligence per year increases at the rate of 0.862. The multi-authorship pattern in the study is found high and the average number of authors per paper is 3.31. Collaborative Index is noted to be the highest range in the year 2014 with 3.50. Mean CI during the period of study is 3.24. This is also supported by the mean degree of collaboration at the percentage of 0.83.The mean CC observed is 0.4635. Lotka's Law of authorship productivity is good for application in the field of artificial intelligence literature. The distribution frequency of the authorship follows the exact Lotka's Inverse Law with the exponent á = 2. The modified form of the inverse square law, i.e., Inverse Power Law with á and C parameters as 2.84 and 0.8083 for artificial intelligence literature is applicable and appears to provide a good fit. Relative Growth Rate [$Rt(P)$] of an article gradually increases from -0.0002 to 1.5405, correspondingly the value of doubling time of the articles $Dt(P)$ decreases from 1.0998 to 0.4499 (2008-2017). At the outset the study reveals the fact that the artificial intelligence literature research study is one of the emerging and blooming fields in the domain of information sciences.

(Batcha, Dar, & Ahmad, 2019) presented a scientometric analysis of the journal titled "Cognition" for a period of 20 years from 1999 to 2018. The present study was conducted with an aim to provide a summary of research activity in current journal and characterize its most aspects. The research coverage includes the year wise distribution of articles, authors, institutions, countries and citation analysis of the journal. The analysis showed that 2870 papers were published in journal of Cognition from 1999 to 2018. The study identified top 20 prolific authors, institutions and countries of the journal. Researchers from USA have made the most percentage of contributions.

(Batcha, Dar, & Ahmad, 2020) conducts ascientometric study of the *Modern Language Journal* literature from 1999 to 2018. A total of 2564 items resulted from the publication name using "Modern Language Journal" as the search term was retrieved from the Web of Science Database. Based on the number of publications during the study period, no consistent growth was observed in the research activities pertaining to the journal. The annual distribution of publications, number of authors, institution productivity, country wise publications and Citations are analyzed. Highly productive authors, institutions, and countries are identified. The results reveal that the maximum number of papers 179 is published in the year 1999. It was also observed that Byrnes H is the most productive, contributed 51 publications and Kramsch C is most cited author in the field having 543 global citations. The highest number (38.26%) of publications, contributed from

**Library and Culture: A Scientometric Analysis and Visualization of Research Trends**

USA and the foremost productive establishment was University of Iowa.

(Ahmad, Batcha, & Dar, 2020) studied the Brain and Language journal which is an interdisciplinary journal, publishes articles that explicate the complex relationships among language, brain, and behaviour and is one such journal which is concerned with investigating the neural correlates of Language. The study aims at mapping the structure of the *Brain and Language* journal. The journal looks into the intrinsic relationship between language and brain. The study demonstrates and elaborates on the various aspects of the Journal, such as its chronology wise total papers, most productive authors, citations, average citation per paper, institution and country wise distribution of publications for a period of 20 years.

## OBJECTIVES

The main objective is to acquire the sciento-graphical mapping of 890 articles published under the topic Library and Culture on Web of Science database during 2010-2019 and the specific objectives are to identify and carry out the following factors:

- Depict the growth of literature in the field of Library and Culture.
- Identify the prolific authors in the Library and Culture field.
- Find out the highly productivity affiliated institutions.
- Find out the highly Contributing Journals.
- Analyze country-wise contributions of the publications.

## METHODOLOGY

All publications on "Library and Culture" in topic were downloaded from Web of Science citation database. The data was exported and processed in Histcite and MS Excel to find out the contribution of Authors, Institutions, Journals, Countries, and Citations in the field of Library and Culture research during years 2010 – 2019. Totally 890 data sets were collected for the study. The year of publication, citations, and authors were analyzed and displayed in tables and scientographs using Histcite and VOSviewer respectively. The global citation scores and local citation scores were examined to identify the pattern of research contribution on Library and Culture.

## DISCUSSION AND RESULT

### Evaluate the Annual Output of Publications

The table I reveals that the numbers of research documents published from 2010 to 2019 are gradually increasing. According to the publication output from the table I the year wise distribution of research documents, 2019 has the highest number of research documents 124 (13.93%) with 1(1.69%) of total local citation score and 72 (3.46%) of total global citation score values and being prominent among the 10 years output and it stood in first rank position. The year 2017 has 114 (12.81%) research documents and it stood in second position with 4 (6.78%) of total local citation score and 394 (10.33%) of total global citation score were scaled. It is followed by the year 2018 with 104 (11.69 %) of records and it stood in third rank position along with 0 (0.00%) of total local citation score and 132 (13.46%) of total global citation score measured. The year 2014 has 94 (10.56%) research documents and it stood in fourth position with 8 (13.53%) of total local citation score and 641 (16.80%) of total global citation score were scaled. It is noticed that the increase in publications may not create impact on citation score yet the quality matters on total local citation scores and on total global citation scores

**Table1.** *Annual Distribution of Publications and Citations*

| S. No. | Year | Records | % | TLCS | % | TGCS | % |
|---|---|---|---|---|---|---|---|
| 1 | 2010 | 73 | 8.20 | 9 | 15.25 | 405 | 10.62 |
| 2 | 2011 | 64 | 7.19 | 4 | 6.78 | 487 | 12.77 |
| 3 | 2012 | 63 | 7.08 | 5 | 8.47 | 567 | 14.86 |
| 4 | 2013 | 73 | 8.20 | 11 | 18.64 | 387 | 10.14 |
| 5 | 2014 | 94 | 10.56 | 8 | 13.56 | 641 | 16.80 |
| 6 | 2015 | 91 | 10.22 | 11 | 18.64 | 390 | 10.22 |
| 7 | 2016 | 90 | 10.11 | 6 | 10.17 | 340 | 8.91 |
| 8 | 2017 | 114 | 12.81 | 4 | 6.78 | 394 | 10.33 |
| 9 | 2018 | 104 | 11.69 | 0 | 0.00 | 132 | 3.46 |
| 10 | 2019 | 124 | 13.93 | 1 | 1.69 | 72 | 1.89 |
|  |  | 890 | 100.00 | 59 | 100.00 | 3815 | 100.00 |

**Library and Culture: A Scientometric Analysis and Visualization of Research Trends**

### Analysis of the Publication Output of Top 10 Authors

Table II and figure 1 displays the ranking of authors of research articles. In the rank analysis the authors who have published 3 or more than 3 articles are considered into account to avoid a long list. It was observed that there are total 1823 authors for 890 records and it shows the top 10 most productive authors during 2010-2019. Lo p published 4 (0.45%) articles with20 TGCS articles, followed by Ho KKW 3 (0.34%) with 20 TGCS articles, Collins M (0.34%) with 11 TGCS articles. The data set clearly depicts that no matter how many publications that an author brings out yet the quality publications alone shows impact in the form of total global citations score.

**Table2.** *Publication output of Top10 Authors and Citation Scores*

| S. No. | Author | Records | % | TLCS | TGCS |
|---|---|---|---|---|---|
| 1 | Lo P | 4 | 0.45 | 1 | 20 |
| 2 | Bressan V | 3 | 0.34 | 0 | 5 |
| 3 | Bruce C | 3 | 0.34 | 2 | 9 |
| 4 | Chidambaranathan K | 3 | 0.34 | 0 | 10 |
| 5 | Collins M | 3 | 0.34 | 1 | 11 |
| 6 | Gonzalez ME | 3 | 0.34 | 0 | 6 |
| 7 | Ho KKW | 3 | 0.34 | 1 | 20 |
| 8 | Johnson IM | 3 | 0.34 | 0 | 0 |
| 9 | Van Schaik S | 3 | 0.34 | 0 | 6 |
| 10 | Zainab AN | 3 | 0.34 | 0 | 5 |

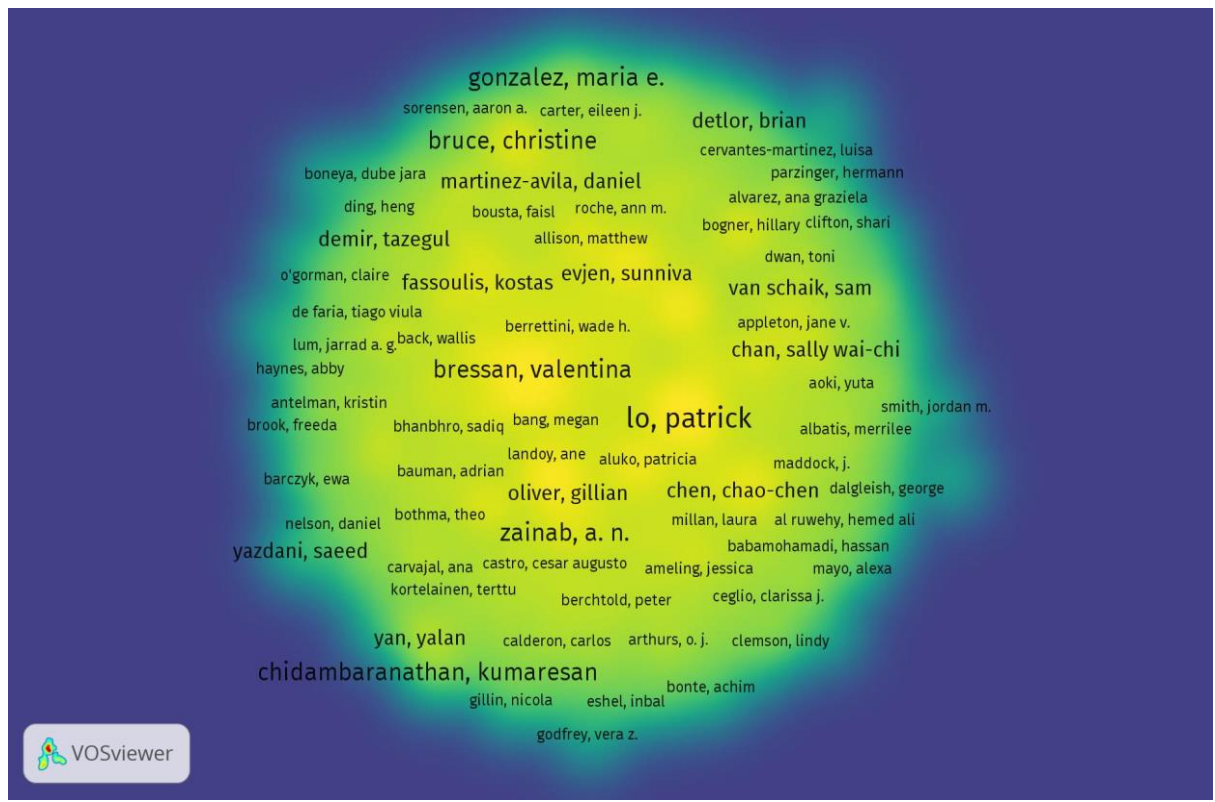

**Figure1.** *Publication output of Top Authors*

### Analysis of the Publication Output of Top 10 Journals

Table III and figure 2 displays the publication output of the top ten journals by number of papers and Journal of Academic Librarianship attained 1st rank among the top ten Journals under consideration with its total global citation score 138. In all 461 journals contributed in research during 2010 and 2019. The journals that rank between 2nd and 10th position is Australian Library Journal, LIBRI, Journal of Librarian and Information Science, College and Research Libraries, Professional De La Information, Information Research and Information Electronic Journal, Journal of Documentation and Electronic Library. We have found by using this journal mapping analysis that there are nodes with clarity of linking between each node, which indicates

**Library and Culture: A Scientometric Analysis and Visualization of Research Trends**

that there are journals linking and associated with other associated journals. It could be identified that the journal wise analysis the following journals: Journal of Academic Librarianship, Australian Library Journal, LIBRI, Journal of Librarian and Information Science, College and Research Libraries, Professional De La Information, Information Research and Information Electronic Journal, Journal of Documentation and Electronic Library were identified the most productive journals based on the number of research papers published.

**Table3.** *Publication output of Top10 Journals and Citation Scores*

| S. No. | Journals | Records | % | TLCS | TGCS |
|---|---|---|---|---|---|
| 1 | Journal of Academic Librarianship | 29 | 3.26 | 11 | 138 |
| 2 | Australian Library Journal | 21 | 2.36 | 6 | 50 |
| 3 | Libri | 19 | 2.13 | 1 | 43 |
| 4 | Library Trends | 17 | 1.91 | 2 | 38 |
| 5 | Journal of Librarianship and Information Science | 15 | 1.69 | 2 | 29 |
| 6 | College & Research Libraries | 14 | 1.57 | 7 | 76 |
| 7 | Profesional De La Informacion | 14 | 1.57 | 0 | 28 |
| 8 | Information Research-An International Electronic Journal | 13 | 1.46 | 0 | 23 |
| 9 | Journal of Documentation | 12 | 1.35 | 4 | 115 |
| 10 | Electronic Library | 11 | 1.24 | 0 | 47 |

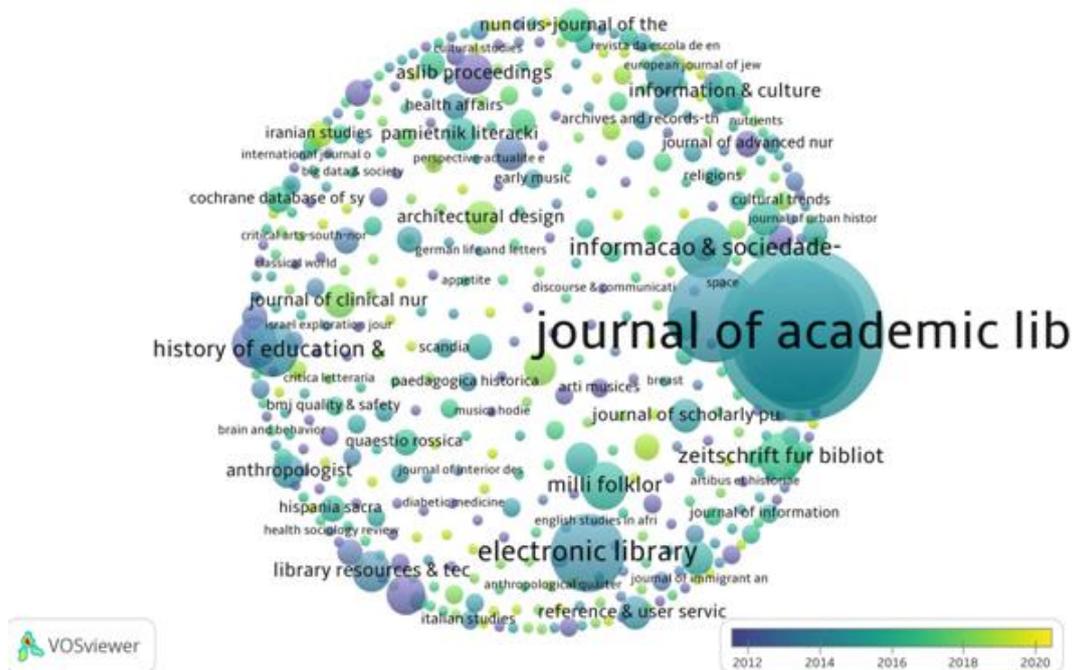

**Figure2.** *Publication output of Top Journals*

### Analysis of the Publication Output of Top 10 Institutions

The individualities of 10 most productive institutions were analyzed in this part, institutions which published more than 2 publications have considered as highly productive institutions. Table IV summarizes articles, the global citation score, local citation score and average citation per paper of the publications of these institutions. In total, 940 institutions, including 1431 subdivisions published 890 research papers during 2010 – 2019. The top most ten prolific institutions involved in this research have published 2 and more research articles. The mean average is 0.95 research articles per institution. Out of 940 institutions, top 10 institutions published 38 (4.27%) research papers and the rest of the institution published 852 (95.73%) research papers respectively. Based on the number of published research records the institutions are ranked.

**Table4.** *Ranking of Institutions and their Research Performance*

| S. No. | Institutions | Records | % | TGCS | ACPP |
|---|---|---|---|---|---|
| 1 | Wuhan Univ, Sch Informat Management | 6 | 0.67 | 15 | 2.5 |
| 2 | Univ Tsukuba, Fac Lib Informat & Media Sci | 5 | 0.56 | 20 | 4 |

| 3 | Univ Cambridge | 4 | 0.45 | 1 | 0.25 |
| 4 | Univ Nacl Autonoma Mexico, Inst Invest Bibliotecol & Informac | 4 | 0.45 | 1 | 0.25 |
| 5 | Victoria Univ Wellington, Sch Informat Management | 4 | 0.45 | 11 | 2.75 |
| 6 | British Lib | 3 | 0.34 | 10 | 3.33 |
| 7 | Penn State Univ | 3 | 0.34 | 16 | 5.33 |
| 8 | RMIT Univ, Melbourne | 3 | 0.34 | 6 | 2.00 |
| 9 | Univ Autonoma Barcelona | 3 | 0.34 | 19 | 6.33 |
| 10 | Univ Guam, Sch Business & Publ Adm | 3 | 0.34 | 20 | 6.67 |

**Analysis of the Publication Output of Top 10 Countries**

Table V and figure 3 displays the publication output of the top ten countries by number of papers and USA acquired 1st rank among the top ten countries under consideration with its total global citation score 708. In all 67 countries contributed in research during 2010 and 2019. The countries that rank between 2nd and 10th position are UK, Australia, Spain, Canada, Italy, People Republic of China, Brazil, Turkey, and Germany have most number of research papers. We have found by using this country mapping analysis that there are nodes with clarity of linking between each node, which indicates that there are countries linking and associated with other associated countries. It could be identified that the country wise analysis the following countries, USA, UK, Australia, Spain, Canada, Italy, People Republic of China, Brazil, Turkey, Germany were identified the most productive countries based on the number of research papers published.

**Table 5.** *Distribution of the Publication Output of Top 10 Countries*

| S. No. | Country | Records | % | TLCS | TGCS |
|---|---|---|---|---|---|
| 1 | USA | 244 | 27.42 | 20 | 1030 |
| 2 | UK | 118 | 13.26 | 6 | 752 |
| 3 | Australia | 76 | 8.54 | 14 | 501 |
| 4 | Spain | 39 | 4.38 | 2 | 257 |
| 5 | Canada | 32 | 3.60 | 5 | 367 |
| 6 | Italy | 30 | 3.37 | 1 | 149 |
| 7 | Peoples R China | 30 | 3.37 | 1 | 92 |
| 8 | Brazil | 25 | 2.81 | 1 | 24 |
| 9 | Turkey | 22 | 2.47 | 3 | 39 |
| 10 | Germany | 17 | 1.91 | 0 | 56 |

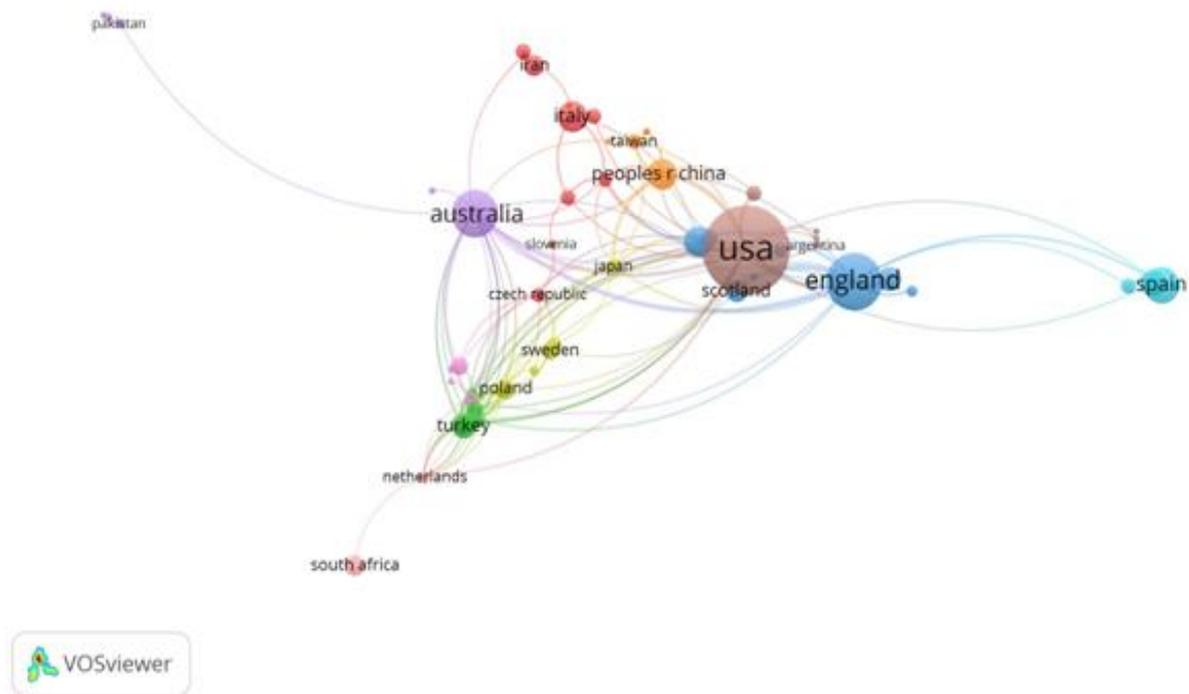

**Figure 3.** *Publication output of Top Countries*

## CONCLUSION

The number of papers published in Web of Science on "Library and Culture" has gradually increased during 2010–2019 and the study has shown that 890 research documents have been published in Web of Science database during the period. It could be identified that the author wise analysis the following authors:Lo P, Bressan V, Brce C and Chidambaranthan K were acknowledged the most prolific authors based on the number of research papers contributed. It could be identified that the institutions wise analysis the following institutions: Wuhan Uni Informant Management, Uni Tsukuba, Fac Lib Informant & Media Sci, and Uni Cambridge were acknowledged the most prolific institutions based on the number of research papers output they published. It could be identified that the journal wise analysis the following journals: Journal of Academic Librarianship, Australian Library Journal, LIBRI, Journal of Librarian and Information Science, College and Research Libraries, Professional De La Information, Information Research and Information Electronic Journal were identified the most productive journals based on the number of research papers published. It could be identified that the country wise analysis the following countries: USA, UK, Australia, Spain, Canada, Italy, People Republic of China, and Germany were identified the most productive countries based on the number of research papers published.